\documentclass{elsart}

\usepackage{natbib}
\usepackage{graphics}
\usepackage{graphicx}
\usepackage{epsfig}
\usepackage{amssymb}

\newcommand{\beq}{\begin{equation}}
\newcommand{\eeq}{\end{equation}}
\newcommand{\beqa}{\begin{eqnarray}}
\newcommand{\eeqa}{\end{eqnarray}}
\newcommand{\bit}{\begin{itemize}}
\newcommand{\eit}{\end{itemize}}

\begin{document}

\begin{frontmatter}


\title{A WENO algorithm for radiative transfer with resonant scattering and
 the Wouthuysen-Field Coupling}

\author[label1]{Ishani Roy},
\author[label2]{Jing-Mei Qiu},
\author[label1]{Chi-Wang Shu},
\author[label3]{Li-Zhi Fang}

\address[label1]{Division of
Applied Mathematics, Brown University, Providence, RI 02912}
\address[label2]{Mathematical and Computer Science,
        Colorado School of Mines, Golden, CO 80401}
\address[label3]{Department of Physics, University of Arizona,
Tucson, AZ 85721}

\begin{abstract}

We develop a numerical solver for the integral-differential
equations, which describe the radiative transfer of photon
distribution in the frequency space with resonant scattering of
Ly$\alpha$ photons by hydrogen gas in the early universe. The
time-dependent solutions of this equation is crucial to the
estimation of the effect of the Wouthuysen-Field (WF) coupling in
relation to the 21 cm emission and absorption at the epoch of
reionization. However, the time-dependent solutions of this equation
have not yet been well performed.  The resonant scattering leads to
the photon distribution in the frequency space to be piecewise
smooth containing sharp changes. The weighted essentially
nonoscillatory (WENO) scheme is suitable to handle this problem, as
this algorithm has been found to be highly stable and robust for
solving Boltzmann equation. We test this numerical solver by 1.) the
analytic solutions of the evolution of the photon distribution in
rest background; 2.) the analytic solution in expanding background,
but without resonant scattering; 3.) the formation of local
Boltzmann distribution around the resonant frequency with the
temperature to be the same as that of atom for recoil. We find that
the evolution of the photon distribution due to resonant scattering
with and without recoil generally undergoes three phases. First, the
profile of the photon distribution is similar to the initial one.
Second, an extremely flat plateau (without recoil) or local
Boltzmann distribution (with recoil) form around the resonant
frequency, and the width and height of the flat plateau or local
Boltzmann distribution increase with time. Finally, the distribution
around the resonant frequency is saturated when the photons from the
source is balanced by the redshift of the expansion. This result
indicates that the onset of the W-F coupling should not be
determined by the third phase, but by the time scale of the second
phase. We found that the time scale of the W-F coupling is equal to
about a few hundreds of the mean free flight time of photons with
resonant frequency, and it basically is independent of the Sobolev
parameter if this parameter is much less than 1.

\end{abstract}


\end{frontmatter}

\section{Introduction}

It is generally believed that detecting redshifted 21 centimeter
signals from early universe is one of the next frontiers in
observational cosmology, because it would be able to provide the
information of the first generation of light sources in the cosmic
dark ages. Many studies have been done on the 21 cm emission and
absorption from the halo of individual first stars (Chuzhoy et al.
2006, Cen 2006, Liu et al. 2007). A common conclusion of these works
is that the configurations of the 21 cm emission and absorption
regions is strongly time-dependent. The reason is simple. The
necessary conditions of 21 cm emission and absorption are that 1.
the fraction of neutral hydrogen (HI) is still high; 2, Ly$\alpha$
photons are available for the Wouthuysen-Field (W-F) coupling.
The region of 21 cm emission around first stars is then a thin shell
just outside the I-front (ionization-front). Hence the 21 cm emission
shell should move with a speed higher than the speed of the I-front
$v_f$, which is rather high, even comparable to the speed of light.
Therefore, the time scale of the formation and evolution of the
regions of 21 cm signal can roughly be estimated by $d/v_f$, $d$ being the
thickness of the 21 cm emission and absorption shell. This time
scale is found to be of the order of 1 Myr and less (Cen 2006,
Liu et al. 2007).

In all of the above-mentioned works, the spin temperature is
calculated with the assumption that the Wouthuysen-Field (W-F)
coupling (Wouthuysen, 1952; Field, 1958, 1959) is effective. That
is, the color temperature of photons around Ly$\alpha$ frequency is
assumed to be the same as the kinetic temperature of hydrogen gas.
The W-F coupling locks the color temperature of Ly$\alpha$ photon to
the kinetic temperature of hydrogen gas, and then links the internal
(spin-) degree of freedom with the kinetic temperature of gas.
Obviously, the W-F coupling would be available
if the time scale of the onset of the W-F coupling is less than the
time scale of the formation and evolution of the 21 cm emission and
absorption shells. However, most calculations on the W-F coupling
are based on the {\it time-independent} Fokker-Planck approximation
(Chen \& Miralda-Escude, 2004; Hirata, 2006; Furlanetto \& Pritchard
2006; Chuzhoy \& Shapiro 2006) of the radiative transfer with resonant
scattering of Ly$\alpha$ photons by hydrogen gas. The time-independent
solution is acceptable only if the time-independent state is approached
in a scale shorter than that of the evolution of the 21 cm signal
regions. Unfortunately, this assumption is not obvious. Time-dependent
solutions are necessary.

The W-F coupling is due to the resonant scattering of Ly$\alpha$
photons by hydrogen atoms. It is described by a Boltzmann-like
integro-differential equation of radiative distribution in the phase
space. The W-F coupling is sensitive to the evolution of the photon
distribution in the frequency space. The analytical solutions of
this equation with resr background has revealed that the frequency
distribution of photons under the resonant scattering of Ly$\alpha$
photons by gaseous hydrogen atoms without recoil has two features:
1. an extremely flat top of the photon profiles in the frequency
space; 2. a sharp boundary of the flat top range(Field 1958).
Currently, the numerical results are still far from precise to match
these features (e.g. Meiksin, 2006).

In this paper, we develop a numerical solver with the weighted
essentially nonoscillatory (WENO) scheme. The WENO method has high
order of accuracy and good convergence in capturing discontinuities
as well as to be significantly superior over piecewise smooth
solutions containing discontinuities (Shu 2003). WENO schemes have
been widely used in applications. It is also effective in solving
Boltzmann equations (Carrillo et al. 2003, 2006) and radiative
transfer (Qiu et al. 2006, 2007, 2008). Therefore, one can expect
that the integral-differential equations of resonant scattering can
be properly handled numerically by the WENO scheme.

The paper is organized as follows. Section 2 presents the basic
equations of the resonant scattering of radiation. Section 3 gives
the numerical solver of the WENO scheme. Section 4 presents the
tests of the numerical solver.  The time scale of the W-F coupling
is briefly addressed in Section 5. A discussion and conclusion are
given in Section 6.

\section{Basic equations}

\subsection{Radiative transfer equations with resonant scattering}

Considering a spatially homogeneous and isotropically expanding
infinite medium consisting of neutral hydrogen with temperature $T$,
the kinetics of photons in the frequency space is described by the
radiative transfer equation with resonant scattering (Hummer \&
Rybicki, 1992; Rybicki \& Dell'antonio 1994)
\begin{eqnarray}
\label{eq1}
 & & \frac{\partial J(x, t)}{\partial t}+2HJ(x, t)-
\frac{cH}{v_T}\frac{\partial J(x, t)}{\partial x}=  \nonumber\\
   & & -kc \phi(x)J(x, t)
+kc\int \mathcal{R}(x,x')J(x',t)dx'+C(t)\phi(x)
\end{eqnarray}
where $J$ is the specific intensity in terms of the photon number,
$H(t)=\dot{R}(t)/R(t)$ is the Hubble parameter, $R(t)$ is the cosmic
factor, $v_T=(2k_BT/m)^{1/2}$ is the thermal velocity of hydrogen
atom, the dimensionless frequency $x$ is related to the frequency
$\nu$ and the resonant frequency $\nu_0$ by $x = (\nu-\nu_0)/\Delta
\nu_D$, and $\Delta \nu_D=\nu_0v_T/c$ is the Doppler broadening. The
parameter $k=\chi/\Delta \nu_D$, and the intensity of the resonant
absorption $\chi$ is given by $\chi=\pi e^2n_1f_{12}/m_ec$, where
$n_1$ the number density of neutral hydrogen HI at ground state, and
$f_{12}=0.416$ is the oscillation strength. The cross section at the
line center is
\begin{equation}
\sigma_0=\frac{\pi e^2}{m_ec}f_{12}(\Delta \nu_D)^{-1}.
\end{equation}
In eq.(\ref{eq1}), $C(t)$ is the source of photons with the frequency
distribution $\phi(x)$, which is the Voigt function of the frequency
profile with the center at $\nu_0$, i.e.
\begin{equation}
\label{eq3}
\phi(x)=\frac{a}{\pi^{3/2}}\int^{\infty}_{-\infty} dy \frac {e^{-y^2}}{(x-y)^2+a^2}
\end{equation}
where $a$ is the ratio of the natural to the Doppler broadening.  We have
$a=A_{21}/(8\pi \Delta \nu_D)$, and $A_{21}=6.25\times 10^8$ Hz
is the Einstein spontaneous emission coefficient. $\phi(x)$ is
normalized with $\displaystyle\int \phi(x')dx'=1$. When
$a\rightarrow 0$, we have pure Doppler broadening as
\begin{equation}
\label{eq4}
\phi_D(x) = \frac{1}{\sqrt{\pi}}e^{-x^2} .
\end{equation}

The redistribution function $\mathcal{R}(x,x')$ gives the
probability of a photon absorbed at the frequency $x'$, and
re-emitted at the frequency $x$. It depends on the details of the
scattering (Henyey 1941; Hummer 1962; Hummer, 1969). If we consider
coherent scattering without recoil, it is
\begin{eqnarray}
\label{eq5}
\lefteqn{ \mathcal{R}(x,x')= } \\ \nonumber
 & \ \ \  & \frac{1}{\pi^{3/2}}\int^{\infty}_{|x-x'|/2}e^{-u^2}
\left [
\tan^{-1}\left(\frac{x_{\min}+u}{a}\right)-\tan^{-1}\left(\frac{x_{\max}-u}{a}\right
)\right ]du
\end{eqnarray}
where $x_{\min}=\min(x, x')$ and $x_{\max}=\max(x,x')$. Obviously,
$\displaystyle\int_{-\infty}^{\infty} \mathcal{R}(x,x')dx'=\phi(x)$. In the case of
$a=0$, i.e. considering only the Doppler broadening, eq.(\ref{eq5}) becomes
\begin{equation}
\label{eq6}
\mathcal{R}(x,x')=\frac{1}{2}{\rm erfc}[{\rm max}(|x|,|x'|)].
\end{equation}
Considering the recoil of atoms, the redistribution function for the
Doppler broadening is (Field, 1959, Basko, 1981)
\begin{equation}
\label{eq7}
\mathcal{R}(x,x')=\frac{1}{2}e^{2bx'+b^2}{\rm erfc}[{\rm
max}(|x+b|,|x'+b|)],
\end{equation}
where parameter $b=h\nu_0/mv_T c= 2.5\times 10^{-4} (10^4/T)^{1/2}$.

\subsection{Re-scaling the equations}

We use the new time variable $\tau$ defined as $\tau= cn_1\sigma_0
t$, which is in units of the mean free flight time of photons at
resonant frequency. The number density of neutral hydrogen atoms
$n_1=f_{\rm HI}n_{\rm H}$, with $f_{\rm HI}$ being the fraction of
neutral hydrogen. For the concordance $\Lambda$CDM model, we have
$n_{\rm H}= 1.88\times 10^{-7} (\Omega_bh^2/0.02)(1+z)^3$ cm$^{-3}$.
The factor 0.75 is from hydrogen abundance. Therefore,
\begin{equation}
\label{eq8} t=0.054 \tau  f^{-1}_{\rm
HI}\left(\frac{T}{10^4}\right)^{1/2}\left (\frac{10}{1+z}\right
)^3\left (\frac{0.022}{\Omega_b h^2}\right ) \hspace{3mm} yrs
\end{equation}
We re-scale the eq.(\ref{eq1}) by the following new variables
\begin{equation}
\label{eq9} J'(x,t)= [R(t)/R_0]^2J, \hspace{5mm} C'(t)=
[R(t)/R_0]^2C.
\end{equation}
Thus, eq.(\ref{eq1}) becomes
\begin{eqnarray}
\label{eq10}
\frac{\partial J'(x, \tau)}{\partial \tau}& =& - \phi(x)J'(x, \tau) \nonumber \\
 & & +\int \mathcal{R}(x,x')J'(x',\tau)dx' + \gamma \frac{\partial J'}{\partial x} +C'(\tau)\phi(x).
\end{eqnarray}
We will use $J$ for $J'$ and $C$ for $C'$ in the equations below. The
parameter $\gamma$  is the so-called Sobolev parameter, $\gamma=
(H/v_T k)=(8\pi H/3A_{12}\lambda^3 n_1)=(H m_e\nu_0/\pi
e^2n_1f_{12})$, where $\lambda$ is the wavelength for Ly$\alpha$
transition. $\gamma$ is simply related to the Gunn-Peterson optical
depth $\tau_{GP}$ by
\begin{equation}
\label{eq11} \gamma^{-1}=\tau_{GP}=4.9\times 10^5 h^{-1}f_{\rm
HI}\left(\frac{0.25}{\Omega_M}\right
)^{1/2}\left(\frac{\Omega_bh^{2}}{0.022}\right )
\left(\frac{1+z}{10}\right )^{3/2}.
\end{equation}
The redshift evolution of $f_{\rm HI}$ is dependent on the reionization
models. Before reionization $f_{\rm HI}\simeq 1$; after reionization
$f_{\rm HI}\simeq 10^{-5}$ in average. Therefore, the parameter $\gamma$ has to
be in the range from 1 ($z\leq 7$) to $10^{-7}$ ($z\geq 10$). For static
background, $\gamma=0$.

\section{Numerical solver:  the WENO scheme}

\subsection{Computational domain and computational mesh}

The computational domain in the case of static background is $x \in [-6, 6]$.
The initial condition is $J(x, 0)=0$. The boundary condition is
\begin{equation}
J(x, \tau) =0, \ {\rm \ at \ } |x| = 6 .
\end{equation}
In the case of expanding background, i.e.$\gamma\neq0$, the computational domain is
bigger than $x \in [-6,6]$ depending on the value of the Sobolev parameter $\gamma$.
The domain $(x_{left},x_{right})$ is chosen such that for the particular value of
$\gamma$ we have
\begin{equation}
J(x_{left}, \tau) \approx 0, \quad J(x_{right}, \tau) \approx 0.
\end{equation}
For example,  the domain is taken to be $x \in [-100,6]$ for the
case of $\gamma=10^{-3}$. Also for different values of $\gamma$ the
solutions reach saturation at different time. For example, we see in
our numerical results that the solution $J(\nu,t)$ of eq.
(\ref{eq9}) reaches saturation at time $\tau = 100$ for $\gamma =
10^{-1}$, and reaches saturation at time $\tau = 10^4$ for $\gamma =
10^{-3}$. The computational domain is discretized into a uniform
mesh in the $x$ direction,
$$x_i=x_{left}+i\Delta x, \hspace{25mm}   i = 0,1,2,\cdots,N_x,$$
where $\Delta x=(x_{right}-x_{left})/N_x$, is the mesh size.
We also denote $J_i^n = J(x_i,\tau^n)$, the approximate solution values at $(x_i,\tau^n)$.

\subsection{Algorithm of the spatial derivative}

To calculate $\frac{\partial J}{\partial x}$, we use the fifth order
WENO method (Jiang \& Shu, 1996). That is,
\begin{equation}
\frac{\partial J(x_i,\tau^n)}{\partial x} \approx \frac{1}{\Delta x}
(\hat{h}_{j+1/2}-\hat{h}_{j-1/2}\\)
\end{equation}
where the numerical flux $\hat{h}_{j+1/2}$ is obtained by the procedure given below.
We use the upwind flux in the fifth order WENO approximation because the wind direction
is fixed (negative).
First we denote
\begin{equation}
h_i = J(x_i,\tau^n),    \hspace{25mm} i = -2, -1,\cdots,N_x+3\\
\end{equation}
where $n$ is fixed. The numerical flux from the WENO procedure is obtained by
\begin{equation}
\hat{h}_{i+1/2}=\omega_1\hat{h}_{i+1/2}^{(1)}+\omega_2\hat{h}_{i+1/2}^{(2)}
+\omega_3\hat{h}_{i+1/2}^{(3)},\\
\end{equation}
where $\hat{h}_{i+1/2}^{(m)}$ are the three third order fluxes on
three different stencils given by
\begin{eqnarray*}
\hat{h}_{i+1/2}^{(1)} &=& -\frac{1}{6}h_{i-1}+\frac{5}{6}h_{i}+\frac{1}{3}h_{i+1},\\
\hat{h}_{i+1/2}^{(2)} &=& \frac{1}{3}h_{i}+\frac{5}{6}h_{i+1}-\frac{1}{6}h_{i+2},\\
\hat{h}_{i+1/2}^{(3)} &=& \frac{11}{6}h_{i+1}-\frac{7}{6}h_{i+2}+\frac{1}{3}h_{i+3},
\end{eqnarray*}
and the nonlinear weights $\omega_m$ are given by,
\begin{equation}
\omega_m =
\frac{\check{\omega}_m}{\displaystyle\sum_{l=1}^3\check{\omega}_l},
\hspace{5mm}
\check{\omega}_l = \frac{\gamma_l}{(\epsilon+\beta_l)^2},\\
\end{equation}
where $\epsilon$ is a parameter to avoid the denominator to become zero
and is taken as $\epsilon = 10^{-8}$. The linear weights $\gamma_l$ are given by
\begin{equation}
\gamma_{1} = \frac{3}{10},\hspace{3mm} \gamma_{2} = \frac{3}{5},
\hspace{3mm} \gamma_{3} = \frac{1}{10},
\end{equation}
and the smoothness indicators $\beta_{l}$ are given by
\begin{eqnarray*}
\beta_1 &=& \frac{13}{12}(h_{i-1}-2h_{i}+h_{i+1})^2 +\frac{1}{4}(h_{i-1}-4h_{i}+3h_{i+1})^2,\\
\beta_2 &=& \frac{13}{12}(h_{i}-2h_{i+1}+h_{i+2})^2 +\frac{1}{4}(h_{i}-h_{i+2})^2,\\
\beta_3 &=& \frac{13}{12}(h_{i+1}-2h_{i+2}+h_{i+3})^2 +\frac{1}{4}(3h_{i+1}-4h_{i+2}+h_{i+3})^2.
\end{eqnarray*}

\subsection{High order numerical integration}

The integration of the resonance scattering term is calculated by a fifth order quadrature
formula (Shen et al. 2007)
\begin{equation}
\int_{x_{left}}^{x_{right}}f(x)dx = \Delta x  \sum_{j=0}^{N_x} w_{j}f(x_j) + O(\Delta x^5),\\
\end{equation}
where the weights are defined as,
\begin{eqnarray*}
&&w_0 = \frac{251}{720}, \quad w_1 = \frac{299}{240}, \quad w_2 = \frac{211}{240}, \quad
w_3=\frac{739}{720}, \\
&& w_{N_x - 3} = \frac{739}{720}, \quad w_{N_x-2} = \frac{211}{240}, \quad w_{N_x - 1} =
\frac{299}{240}, w_{N_x} = \frac{251}{720}, \\
\end{eqnarray*}
and  $w_j = 1$ otherwise.
Notice that this numerical integration is very costly and uses most part of the CPU
time.  For the equation with recoil and the redistribution function with $a=0$, we have
used a grouping of the numerical
integration operations at different $x$ locations so that the computational cost
can be reduced to order $O(N)$ rather than $O(N^2)$, where $N$ is the number of
grid points in $x$, without changing mathematically the algorithm and its accuracy.
The $O(N)$ algorithm for numerical integration is given in the appendix which further
highlights the speed and accuracy of the numerical algorithm proposed in this paper.
Unfortunately, this grouping technique does not work for the case $a \neq 0$, hence
the CPU cost for the case with $a \neq 0$ is much larger.

\subsection{Time evolution}

To evolve in time, we use the third-order TVD Runge-Kutta time
discretization (Shu \& Osher, 1988). For systems of ODEs $u_t =
L(u)$, the third order Runge-Kutta method is
\begin{eqnarray*}
u^{(1)} &=& u^n + \Delta \tau L(u^n,\tau^n),\\
u^{(2)} &=& \frac{3}{4}u^n + \frac{1}{4}(u^{(1)}+\Delta \tau L(u^{(1)},\tau^n +
\Delta \tau)),\\
u^{(3)} &=& \frac{1}{3}u^n + \frac{2}{3}(u^{(2)}+\Delta \tau L(u^{(2)},\tau^n +
\frac12 \Delta \tau)).
\end{eqnarray*}

\section{Tests of the WENO solver}

\subsection{Static background}

\begin{figure}[htb]
\centering
\includegraphics[width=7.5cm]{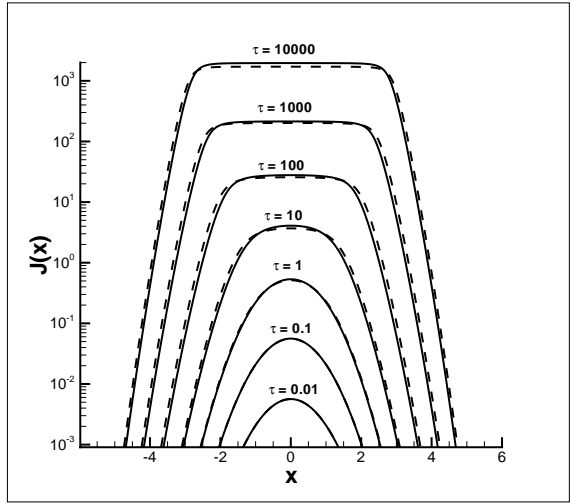}
\caption{Static solutions ($\gamma=0$) for pure Doppler
redistribution ($a=0$) of eq.(\ref{eq10}), in which $C=1$ and
$J(x,0)=0$. The analytical solutions are shown by dashed lines,
while the numerical results are shown by solid lines.} \label{fig1}
\end{figure}

\begin{figure}[htb]
\centerline{
\includegraphics[width=8.0cm]{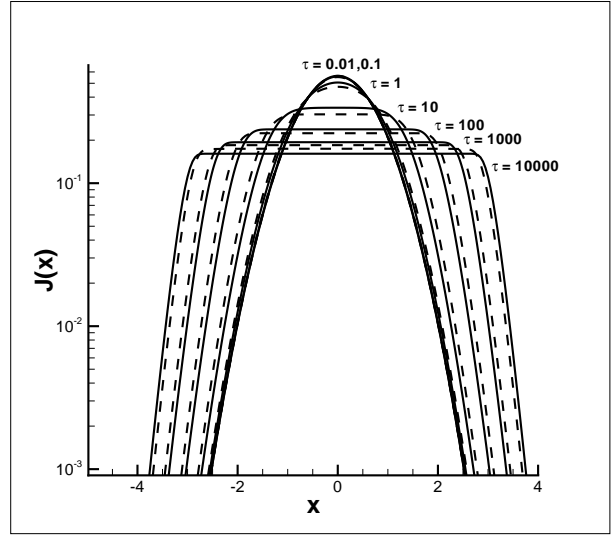}
} \caption{Static solutions ($\gamma=0$) for pure Doppler
redistribution ($a=0$) of eq.(\ref{eq10}), in which $C=0$ and
$J(x,0)=\pi^{-1/2}\exp(-x^2)$. The analytical solutions are shown by
dashed line, while the numerical results are shown by solid lines. }
\label{fig2}
\end{figure}

We first test the WENO solver with two analytical solutions of eq.
(\ref{eq10}) with static background, $H=\gamma=0$ and Doppler
broadening $a=0$ (Field, 1958). The first analytical solution is for
the initial radiative field $J(x,\tau)=0$ and the constant source
$C(\tau)=1$. It is
\begin{eqnarray}
\label{eq19}
\lefteqn{ J(x, \tau)=\pi^{-1/2}[1-\exp(-\tau e^{-x^2})]}\\ \nonumber
  & & \ \hspace{2cm} + \int_{x}^{\infty}e^{w^2}[1-(1+\tau e^{-w^2})
\exp(-\tau e^{-w^2})]{\rm erf}(w)dw.
\end{eqnarray}
The second analytical solution of eq.(\ref{eq10}) is also for $H=\gamma=0$, $a=0$,
but the source $C=0$, while the initial radiative field is
\begin{equation}
\label{eq20}
J(\nu, 0) =\pi^{-1/2}e^{-\nu^2}.
\end{equation}
The solution is
\begin{equation}
\label{eq21}
J(x,\tau)=\pi^{-1/2} e^{-x^2}\exp(-\tau e^{-x^2})+ \tau
\int_{x}^{\infty}e^{-w^2}\exp(-\tau e^{-w^2}){\rm erf}(w)dw.
\end{equation}
The analytical solutions (\ref{eq19}) and (\ref{eq21}) are shown in
Figures~\ref{fig1} and \ref{fig2} respectively. We also plot the
numerical solutions given by our algorithm in same figures. The
numerical results show very small deviation from the analytical
solutions.

A common feature of Figures 1 and 2 is that the originally Doppler
peak at the center of the frequency profile gradually becomes a
flat plateau. The width of the plateau increases with time. This is
because the resonant scattering makes a non-uniform
distribution in the frequency space to a uniform one. It is similar to that
in the physical space, diffusion generally leads to an evolution from non-uniform
distribution to a uniform one. The height of the plateau of Figure 1 is increasing
with time, while in Figure 2 it is decreasing, because for the case $C=1$, the
number of photons increases, while for the case of $C=0$, the total number of photons
is conserved.

\subsection{Expanding background }

\begin{figure}[htb]
\centerline{
\includegraphics[width=2.5in,height=2.3in]{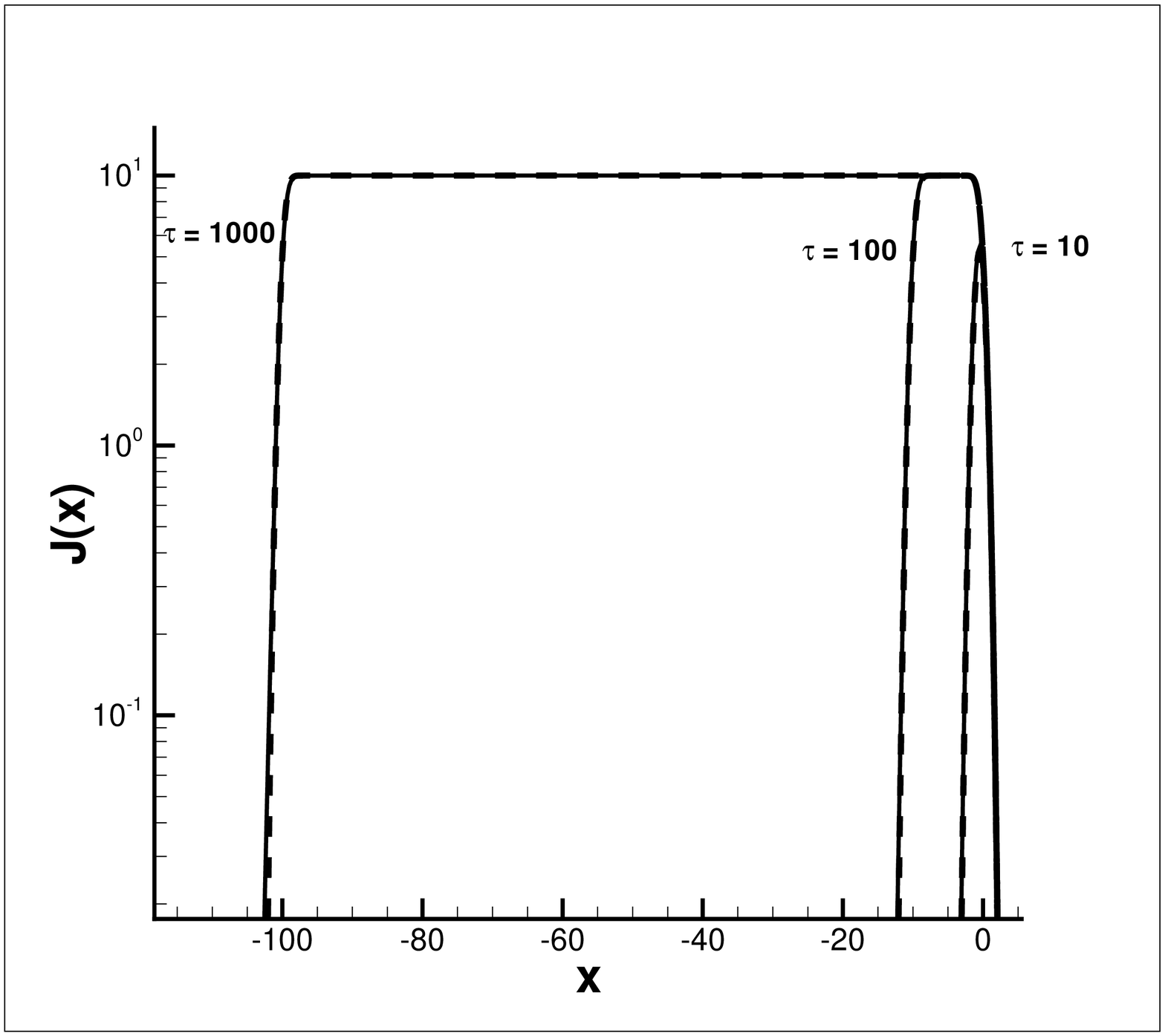}
\includegraphics[width=2.5in,height=2.3in]{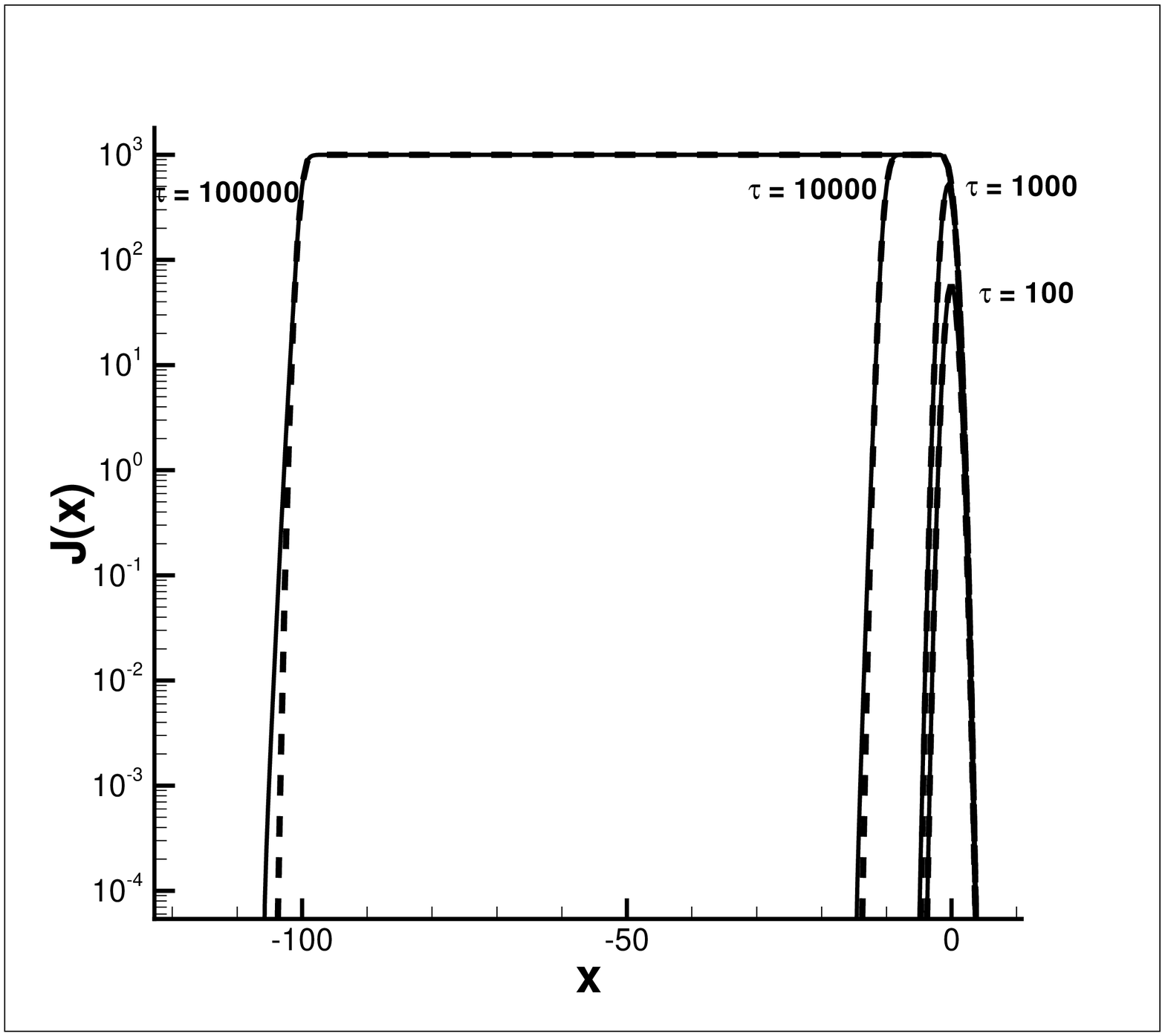}
} \caption{Solutions of $J(x,\tau)$ of eq.(\ref{eq23}) with initial
condition $J(x,0)=0$. The parameter $\gamma$ is taken to be
$10^{-1}$ (left panel) and $10^{-3}$ (right).  The analytical
solutions eq.(24) are shown by dashed lines}. 
\label{fig3}
\end{figure}

\begin{figure}[htb]
\centerline{
\includegraphics[width=2.5in,height=2.3in]{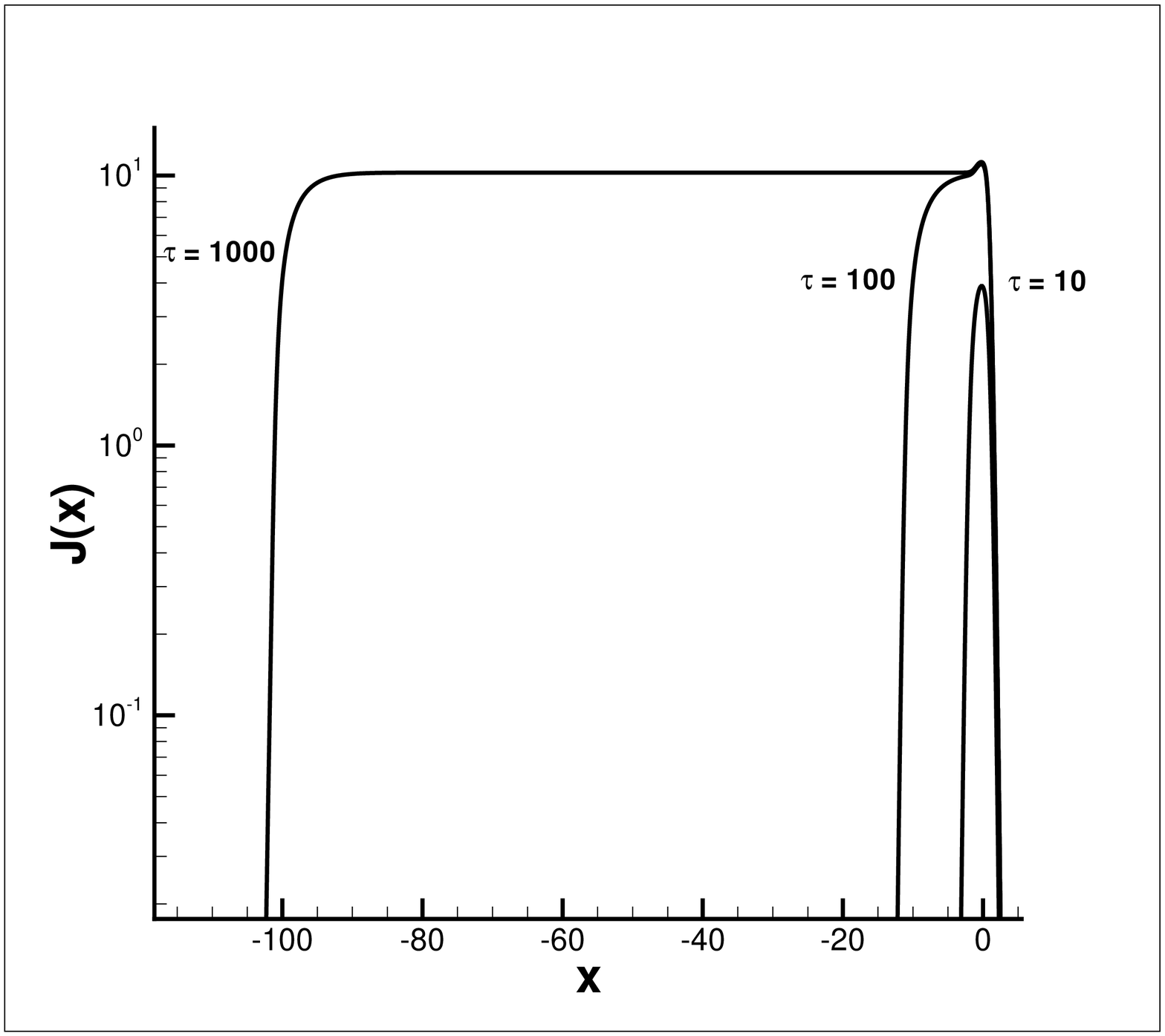}
\includegraphics[width=2.5in,height=2.3in]{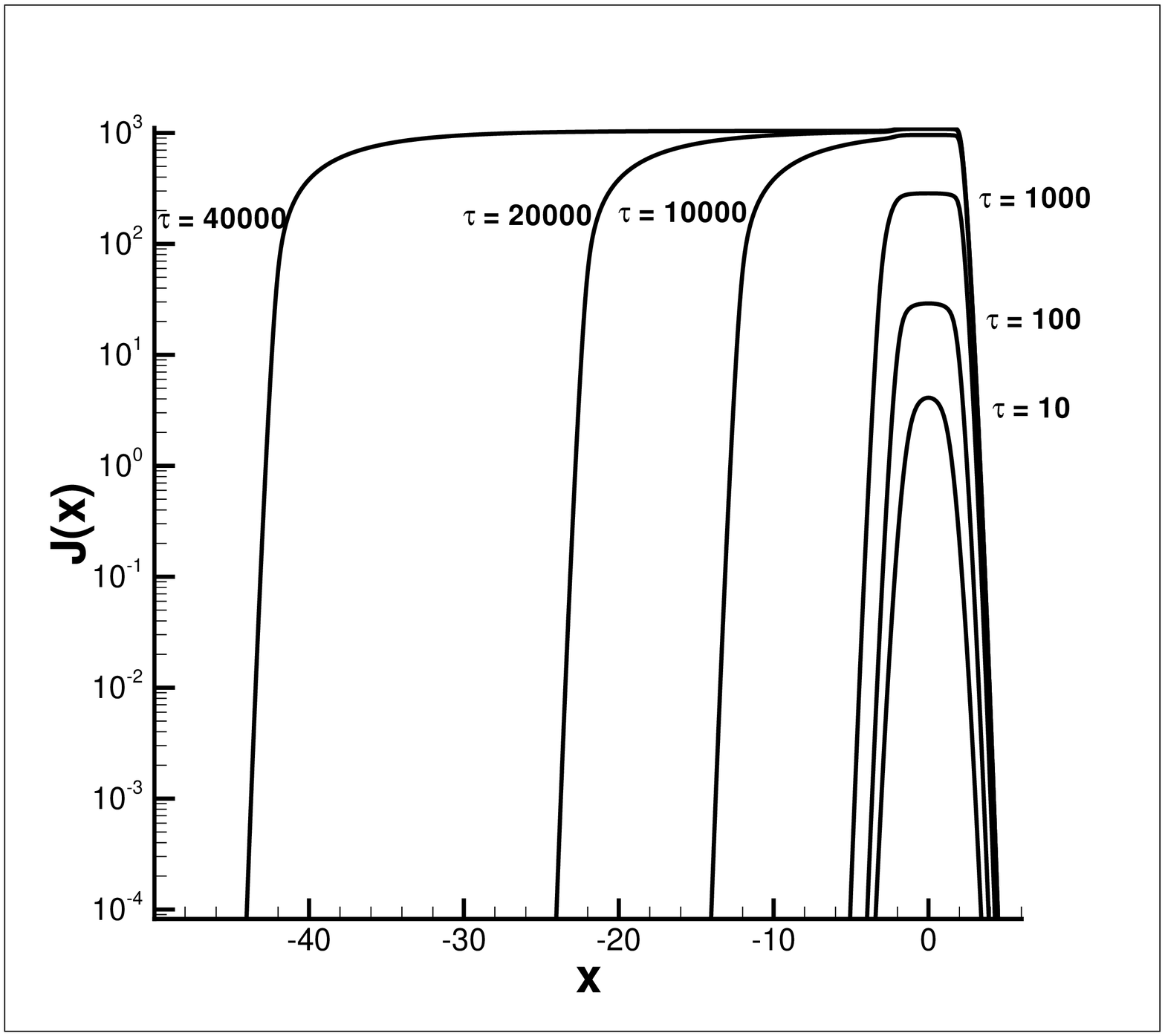}
}
\caption{Solutions of $J(x,\tau)$ of eq.(\ref{eq10}) with initial condition $J(x,0)=0$.
The parameter $\gamma$ is taken to be $10^{-1}$ (left) and $10^{-3}$
(right).}
\label{fig4}
\end{figure}

The second test is given by considering expanding background. Without absorption and
scattering, eq.(10) becomes
\begin{equation}
\label{eq23}
\frac{\partial J}{\partial \tau} =
\gamma \frac{\partial J}{\partial x} +C(\tau)\phi(x).
\end{equation}
If $C$ is $\tau$-independent, the analytic solution of eq.(23) is
(Rybicki \& Dell'antonio, 1994)
\begin{equation}
\label{eq24}
J(x,\tau)=J(x+\gamma \tau, 0)+C\gamma^{-1}[\Phi(x)- \Phi(x+\gamma\tau)]
\end{equation}
where
\begin{equation}
\label{eq25}
\Phi(x)=\int_x^{\infty}\phi(x')dx'
\end{equation}
The meaning of the solution eq.(24) is simple. It just shows that the
redshift of photons in frequency space is described by
\begin{equation}
-x = \gamma \tau .
\end{equation}

We take $\phi(x)=(1/\sqrt{\pi})e^{-x^2}$, and assuming the initial
field $J( x,0)=0$. Figure 3 shows that the numerical results obtained by using the WENO algorithm agree well with the analytic solution of eq.(24).

Figure 4 is the solutions of eq.(10) with parameter $\gamma=10^{-1}$
and $10^{-3}$. In these cases the analytical solutions are not available. We can
see from Figure 3 and 4 that the intensity $J$ stops to
increase and approaches a saturated value when $\tau$ is large. This happens when the number of photons produced from the sources is equal
to that of the redshifted photons.

\subsection{The effect of recoil of atom}

\begin{figure}[htb]
\centerline{
\includegraphics[width=2.8in,height=2.5in]{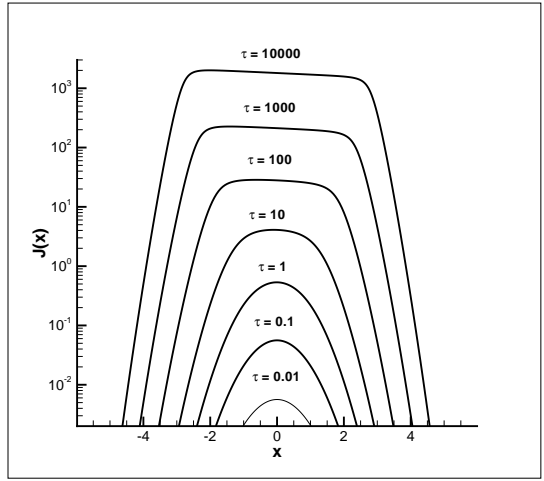}
}
\caption{Solutions of $J(x,\tau)$ of eq.(\ref{eq10}) with redistribution function eq.(7), and
$\gamma=0$.}
\label{fig5}
\end{figure}

The third test is by using the redistribution function with recoil
Eq.(7). Figure 5 gives the solutions of eq.(10) with parameter
$b=0.03$, and the initial condition and source term are taken to be
the same as Figure 1. The results of Figure 5 actually is similar to
Figure 1, but only the center flat plateau of Figure 1 now is
replaced by a sloping plateau. The latter is a local Boltzmann
distribution around the renasont scattering.

Field (1959) has shown that, once the solution $J(x,\tau)$ of eq.(\ref{eq10})
with $b=0$  has a flat plateau in the central region, the effect of
recoil is to make the flat plateau to be a Boltzmann-like distribution, i.e. if the
solution with no recoil redistribution eq.(\ref{eq6}) shows
$J(x,\tau)\simeq J(0,\tau)$, within $|x|< x_f$, the solution of eq.(10) with
redistribution eq.(\ref{eq23}) will be
\begin{equation}
J(0,\tau)e^{-2bx}=J(0,\tau)e^{-h(\nu - \nu_0)/kT}, \hspace{5mm}
|x|< x_f.
\end{equation}
One can test this property with $F(\tau)$ defined as
\begin{equation}
F(\tau)=\log J(0,\tau)-\log(1,\tau)
\end{equation}
If the local Boltzmann distribution within $|x|<1$ is realized at
time $\tau$, $F(\tau)$ should be equal to $2b$ [eq.(27)]. In Figure
6, we present the relation of $F(\tau)$ vs. $\tau$. At $\tau=0$, the
Gaussian sources [eq.(21)] yields $F(0)=1$. Then, $F(\tau)$
approaches to $2b$ at $\tau > 10^2$. These results show that the
WENO solver is reliable and effective to study the time-dependence
of $J$ described by the resonant scattering equation.

Figure 6 also shows the relation $F(\tau)$ vs. $\tau$ for solutions
$b=0.03$ (left panel) and $b=0$ (right panel). The two curves
actually are similar. When the right curve approaches to
$F(\tau)\rightarrow 0$ at time $\tau$, the left curve, at the same
time, approaches to $F(\tau)\rightarrow 2b$. Therefore, it would be
reasonable to time-scale of the formation of local Boltzmann
distribution by the formation of the flat plateau.

\begin{figure}[htb]
\centerline{
\includegraphics[width=2.4in,height=2.2in]{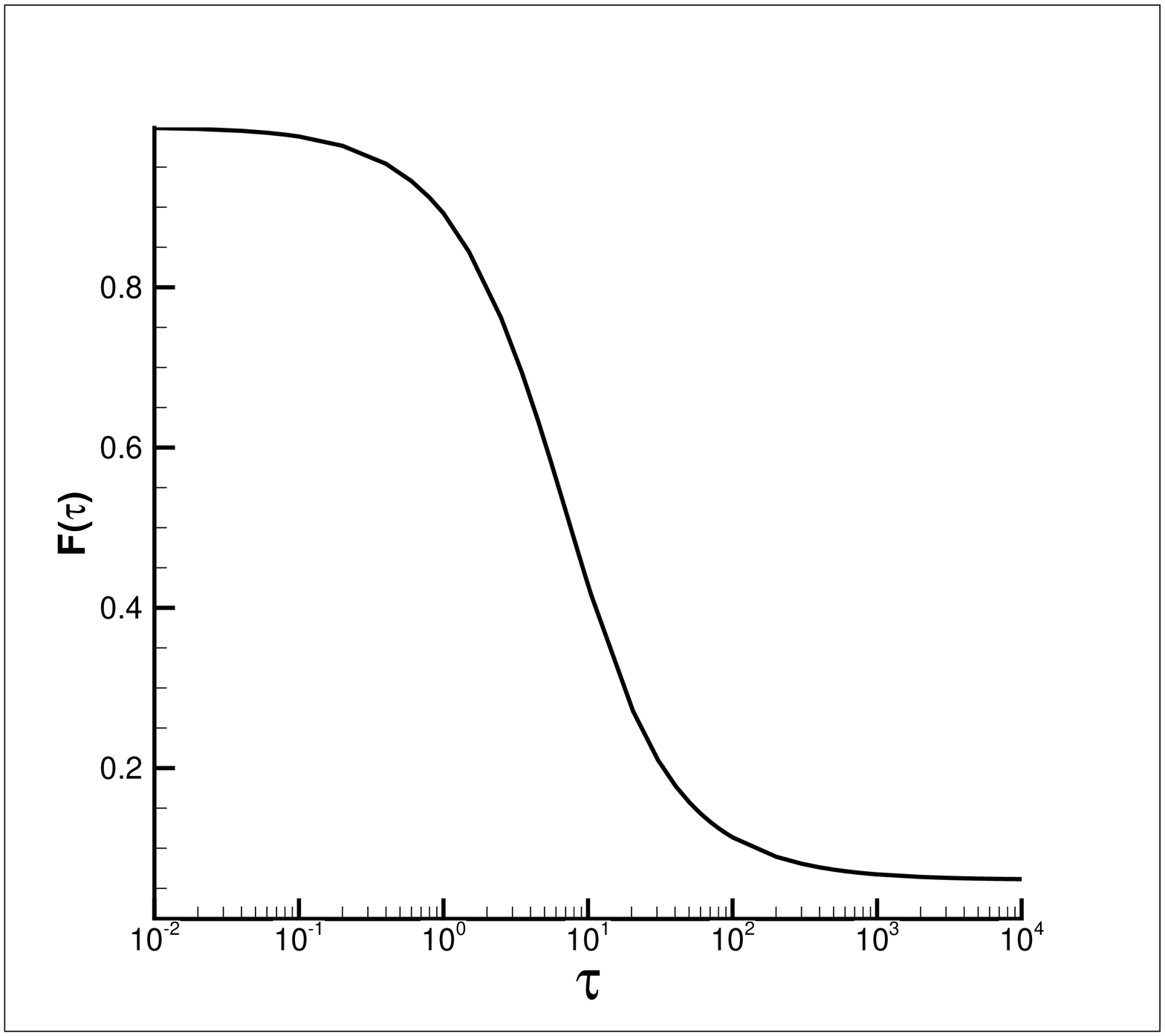}
\includegraphics[width=2.4in,height=2.2in]{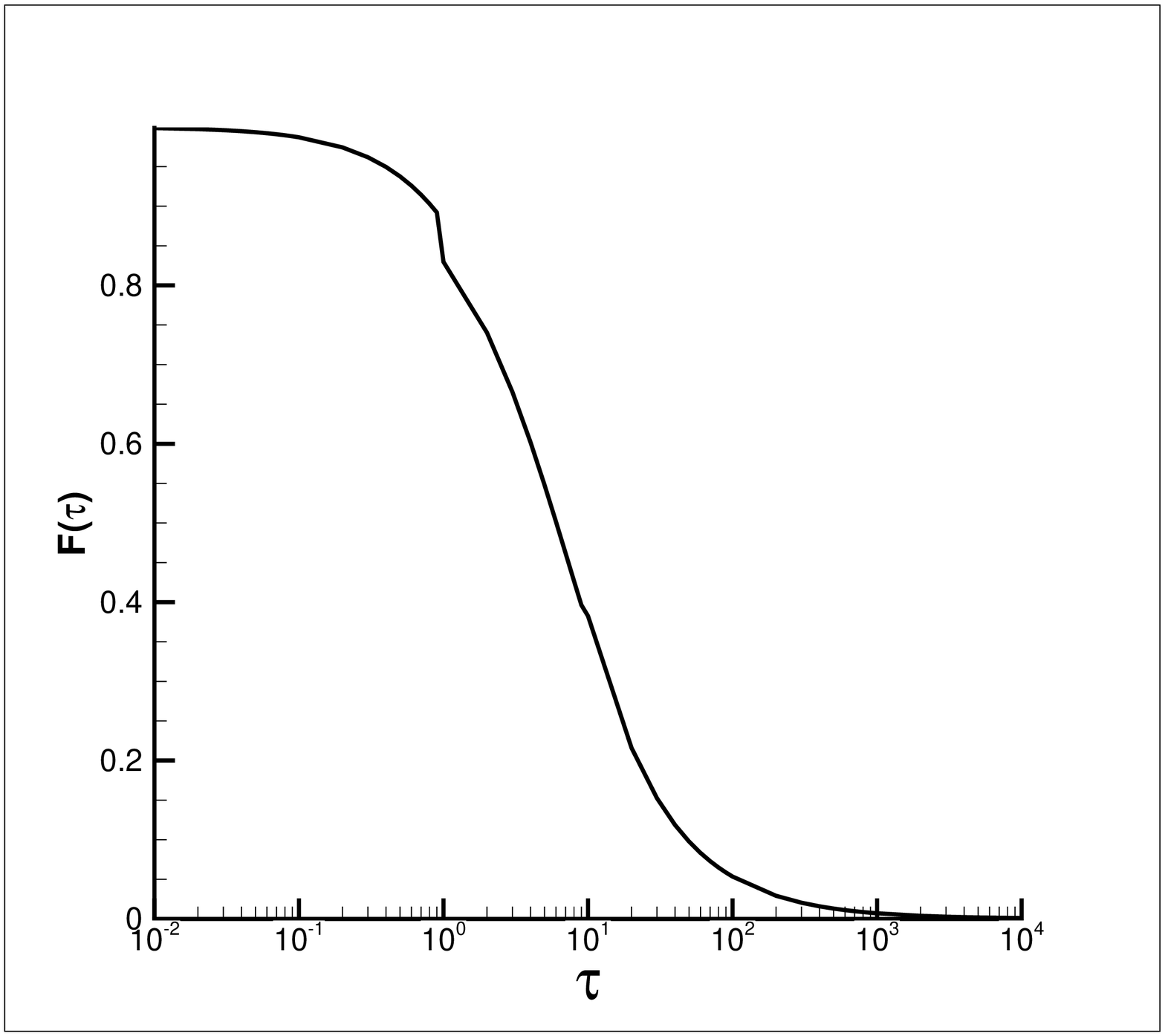}
} \caption{$F(\tau)$ as function of $\tau$ for solutions shown in Figure 5 (left) and
Figure 1 (right)}
\label{fig6}
\end{figure}

\section{The Wouthuysen-Field Coupling}

We now estimate the time scale, $\tau_{\rm WF}$, of the onset of the
W-F coupling. As mentioned in \S 4, in the first phase, $J(x,\tau)$
keeps the initial profile $\phi(x)$ of the photons from source $C$.
In the second phase, the profile is no longer the initial one. A
flat plateau (without recoil) or local Boltzmann distribution (with
recoil) form the central part ($|x |\simeq 0$) due to the resonant
scattering. The width of the flat plateau increases with the time
$\tau$. Finally, in the third phase, the injection of Ly$\alpha$
photons from the source $C$ is balanced by the redshift, the height
of the flat plateau or the local Boltzmann of $J(x,\tau)$ will stop
increasing, and reach to a saturated value. The width of the flat or
sloping plateau on the red side continuously increases.

As mentioned in Section 1, most works on the effect of
the W-F coupling of 21 cm signal are based on the static solution of the
Fokker-Planck approximation of eq.(10) (Chen \& Miralda-Escude, 2004;
Hirata, 2006; Furlanetto \& Pritchard 2006; Chuzhoy \& Shapiro 2006).
Obviously, this solution corresponds to the third phase. Thus,  the time scale
of the W-F coupling is of the order of $\tau_{\rm III}$. Therefore,
this approximation would be reasonable if the time scale of the
formation and evolution of the 21 cm signal regions is larger than
$\tau_{\rm III}$.

From Figures~\ref{fig4} and \ref{fig6}, one can see that the time
scale of the onset of the third phase, $\tau_{\rm III}$, has to be
of the order of $10^2$, and $10^{4}$ for $\gamma=$ $10^{-1}$, and
$10^{-3}$ respectively, i. e. $\tau_{\rm III}$ is roughly equal to a
few to ten of Gunn-Peterson optical depth. On the other hand, we can
see from eq.(\ref{eq11}) that $\gamma=$ $10^{-1}$, and $10^{-3}$
correspond to $f_{\rm HI}=$ $10^{-4}$, and $10^{-2}$ for the source
at redshift $1+z=10$. Thus, for all cases, eq.(\ref{eq8}) yields
that the time scale $t_{\rm III}$ to be of the order of 1 Myr. The
static solution of the Fokker-Planck equation would not be valuable
for the 21 cm problem, if the time scale of the evolution of the 21
cm region is equal to or less than 1 Myr.

Actually the time-independent solution of eq. (\ref{eq10}) would not
be necessary for the 21 cm signal. The relative occupation of the
two hyperfine-structure components of hydrogen ground state depends
only upon the shape of the spectrum near the Ly-alpha frequency,
regardless whether the solution is time-independent, or saturated.
What we need for the W-F coupling is only the frequency distribution
to show a local Boltzmann-like distribution $J(x,\tau)\propto \exp[-
h(\nu-\nu_0)/kT]$ around $\nu_0$, where $T$ is the kinetic
temperature of hydrogen gas. The formation of the Boltzmann-like
distribution is irrelevant to the time scale $\tau_{\rm III}$, but
depends on the onset of the second phase of the $J(x,\tau)$
evolution.

Therefore, $\tau_{\rm WF}$ should be estimated by the time scale of
the onset of the second phase, $\tau_{\rm II}$, not by $\tau_{\rm
III}$. According to the property shown by Figure 6, $\tau_{\rm II}$
can be found by the formation of a flat plateau around the resonant
frequency. For small $\gamma$, $\tau_{\rm II}$ is much less than
$\tau_{\rm III}$. To show this point, we calculate the solution of
eq.(10) with $\gamma=10^{-5}$ and $10^{-7}$ on static background.
The results are given in Figure 7. It shows that that a small flat
plateau has already formed at $\tau=100$. On the other hand,
$\tau_{\rm III}$ will be as large as $\tau\simeq 10^6$ for
$\gamma=10^{-5}$.

\begin{figure}[htb]
\centerline{
\includegraphics[width=2.4in,height=2.1in]{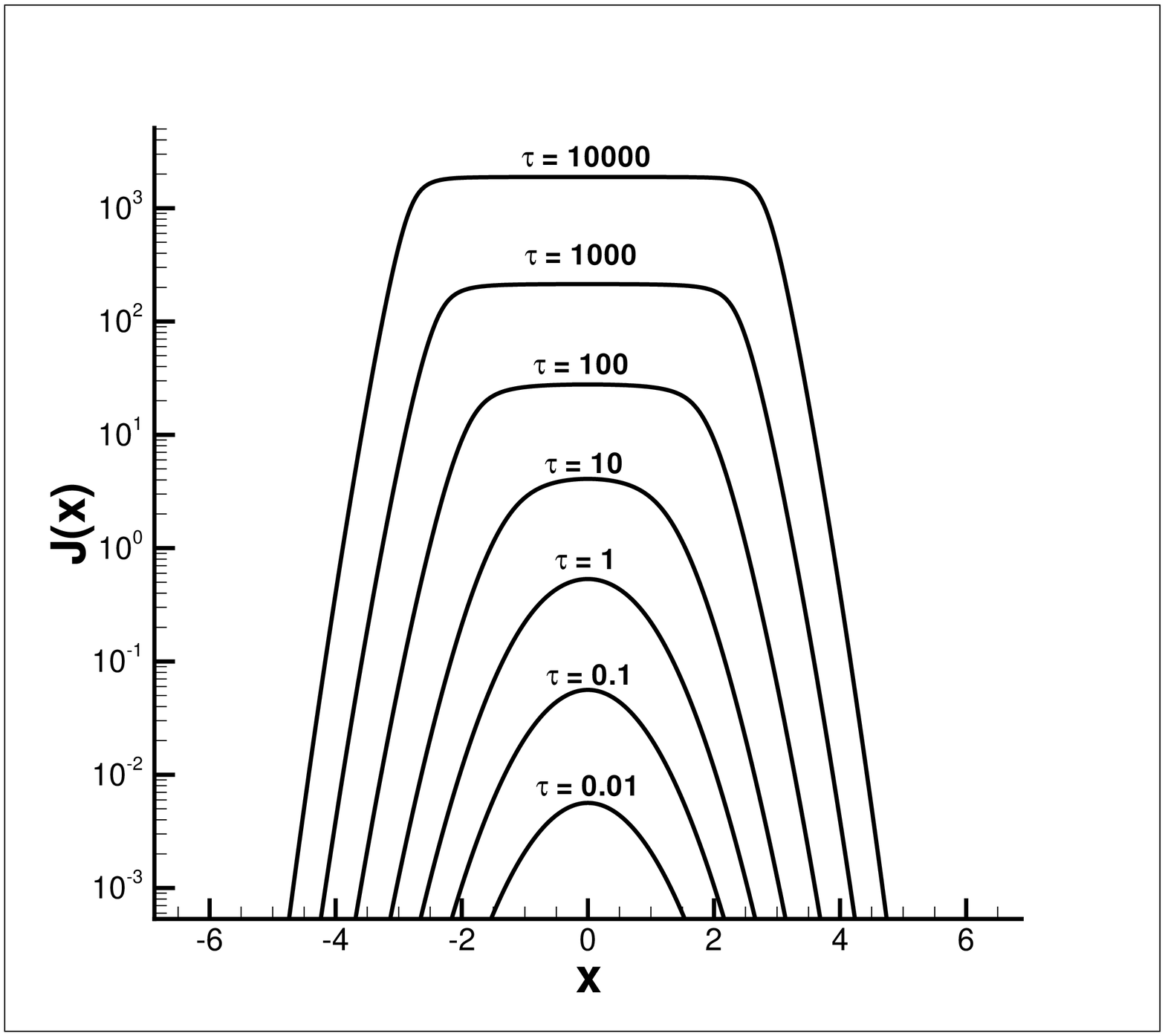}
\includegraphics[width=2.4in,height=2.1in]{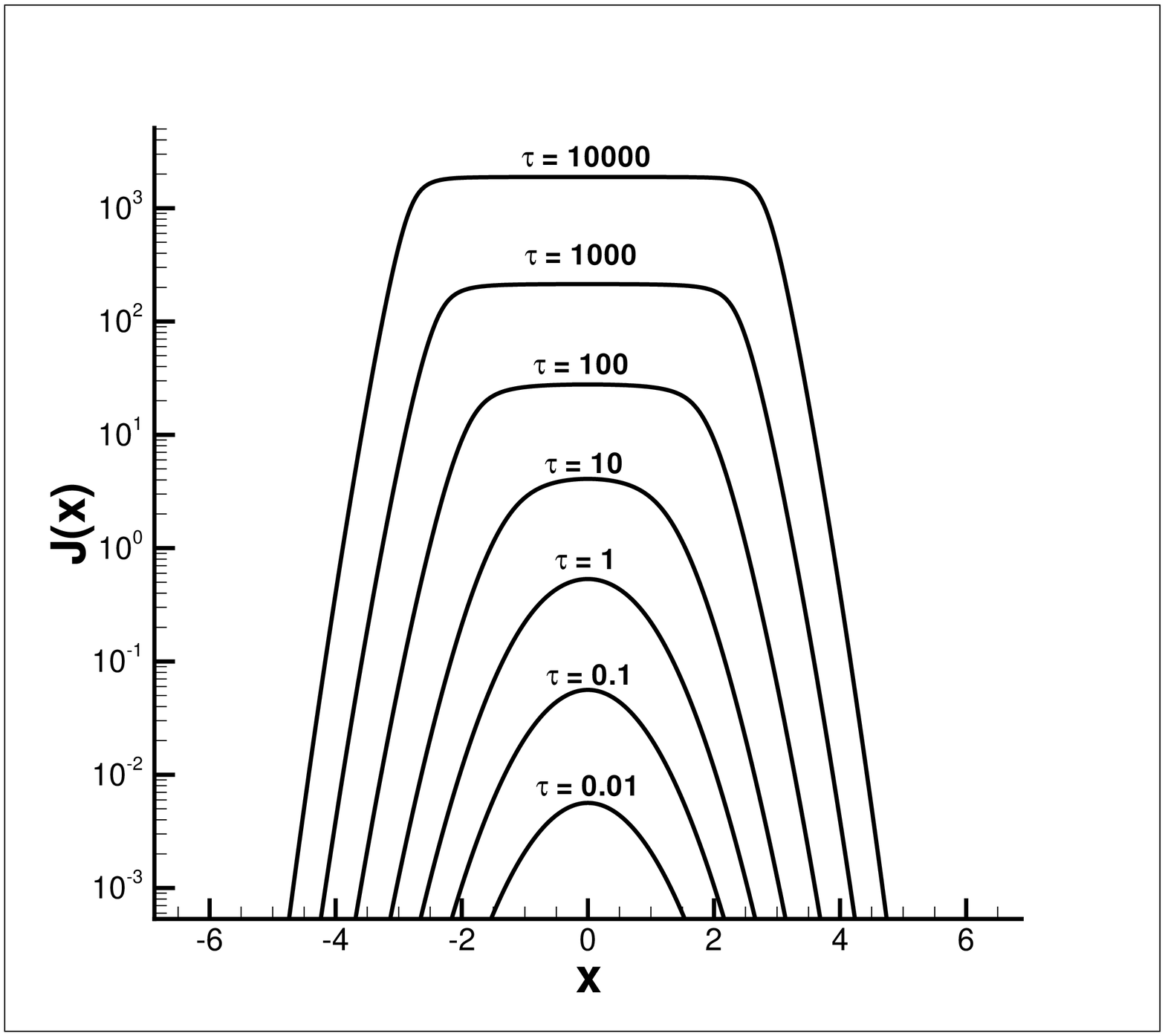}
} \caption{Solutions of $J(x,\tau)$ of eq.(10) with $\gamma=10^{-5}$
(left panel) and 10$^{-7}$ (right panel).}
\label{fig7}
\end{figure}

\begin{figure}[htb]
\centerline{
\includegraphics[width=2.4in,height=2.1in]{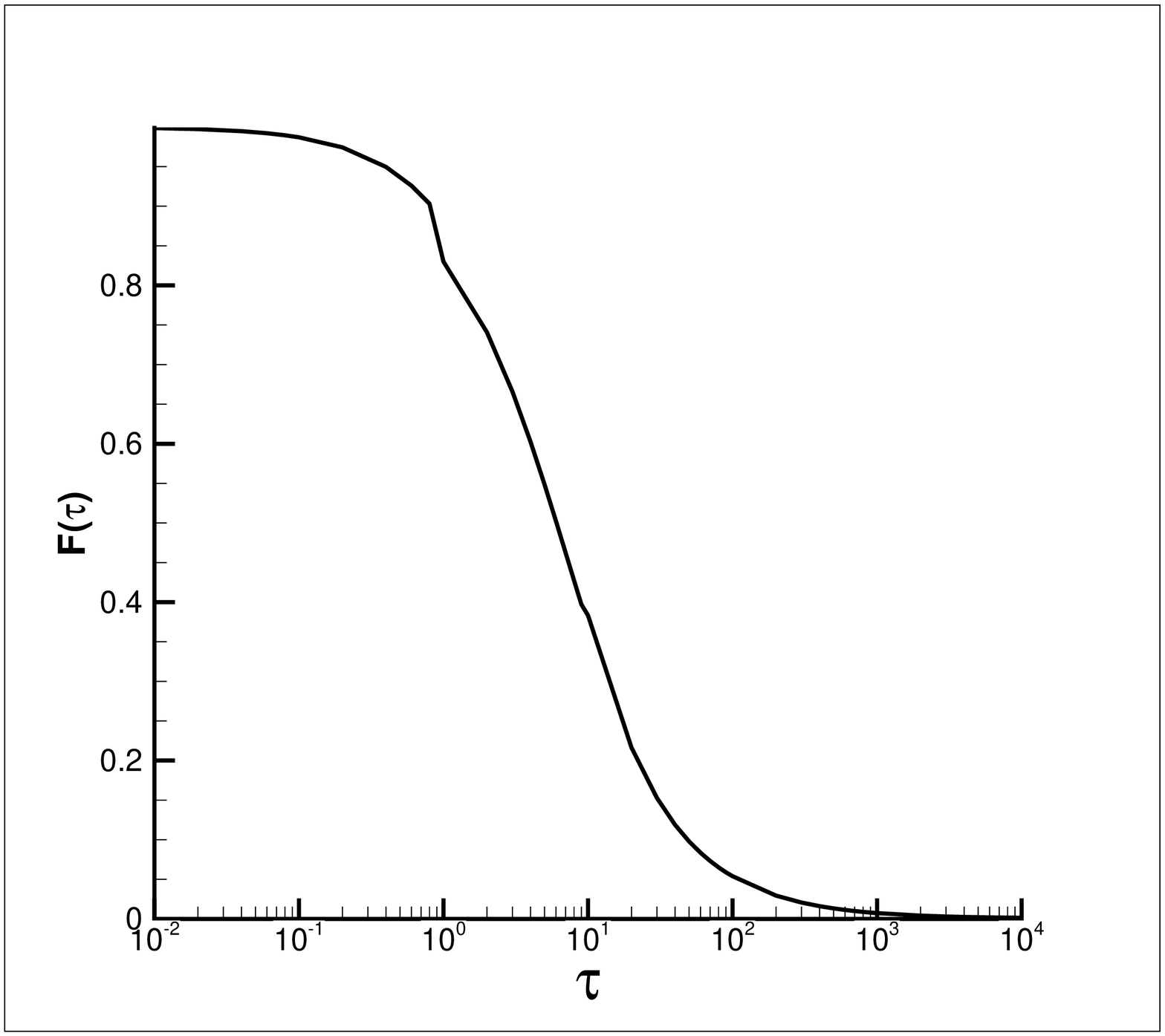}
\includegraphics[width=2.4in,height=2.1in]{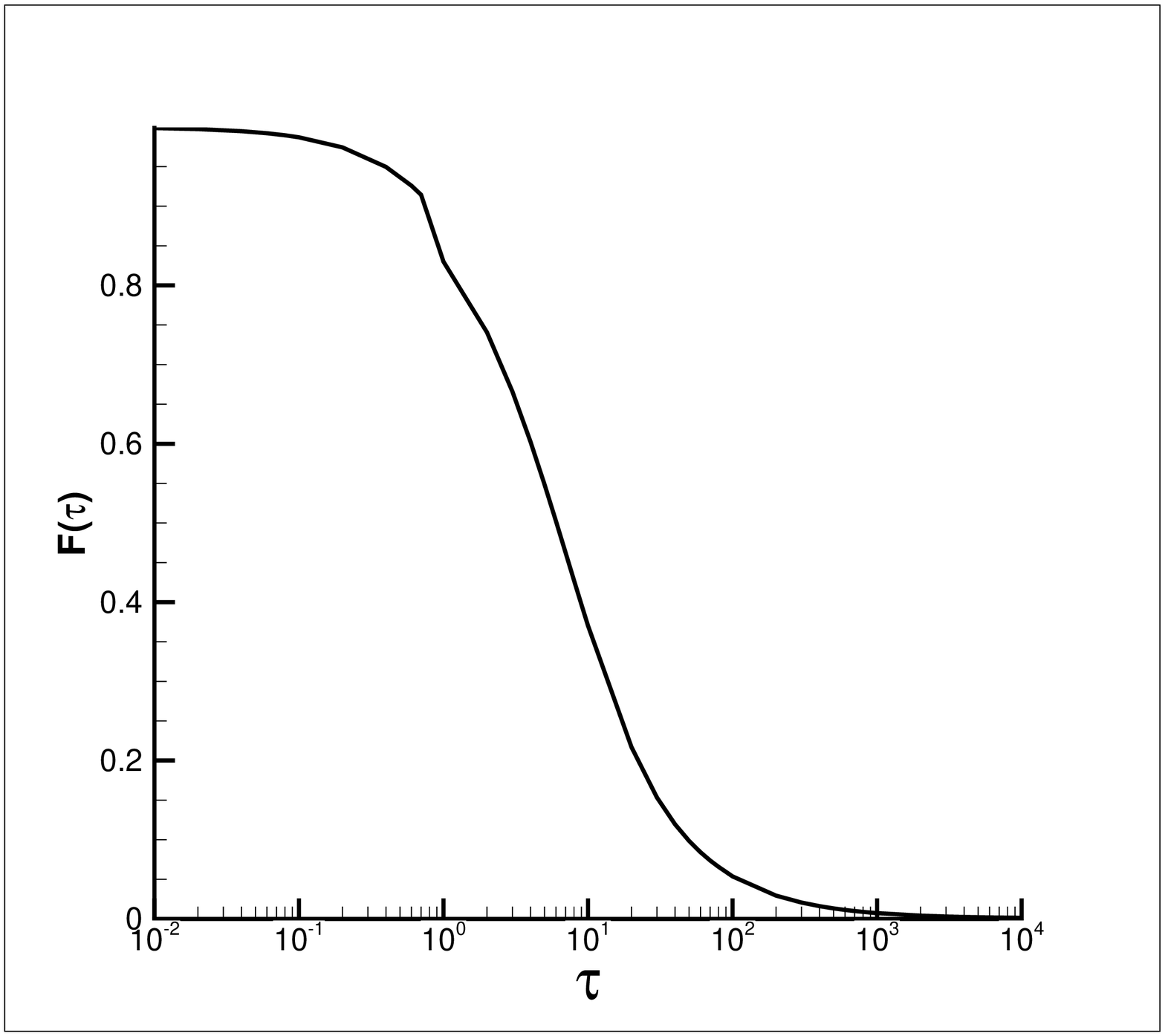}
} 
\caption{$F(\tau)$ as function of $\tau$ for solutions of $a=0$
and $\gamma = 10^{-5}$ (left) and 10$^{-7}$(right). 
} 
\label{fig8}
\end{figure}

In Figure~\ref{fig8}, we show the function $F(\tau)$ for the
parameters  ($a=0$, $\gamma=10^{-5}$), and ($a=0$,
$\gamma=10^{-7}$). It is interesting to observe that the two sets of
$F(\tau)$ are almost identical. That is, the time scale $\tau_{\rm
WF}$ is actually independent of the parameter $\gamma$. For both 
cases of the parameter $\gamma$, $\tau_{\rm WF}$ is equal to
about a few of 10$^2$. In other words, a flat plateau with proper
width can form with a few tens or hundreds of the mean free
flight time of the resonant photons, regardless the expansion of the
universe.

The reason of the $\gamma$-independence of $\tau_{\rm II}$ or
$\tau_{\rm WF}$ can be directly seen from eq.(10). This equation
describes two kinetic processes of approaching statistically steady
state. The first is the resonant scattering, which leads to the
formation of a flat plateau or local Boltzmann distribution around
the resonant frequency. The second is the redshift of photons, which
leads to a statistical equilibrium between the injected photons and
redshifted photons. The steady state due to resonant scattering
corresponds to the second phase, while the steady state due to
redshift corresponds to the third phase. Similar to various
statistical equilibrium or steady state maintained by collision or
scattering, the steady state of resonant scattering can be realized
via a few tens or hundreds of the resonant scattering. On the other
hand, $\gamma$ is a ratio between the time scales of the resonant
scattering and the expansion of the universe. Therefore, when
$\gamma$ is much less than 1, the time scale of the formation of the
flat plateau and local Boltzmann distribution are much shorter than
the expansion of the universe. This point can also be seen with
Figure~\ref{fig6}, which shows that for $\gamma=10^{-5}$ and
10$^{-7}$ the evolutions of $J(x,\tau)$ actually are the same till
$\tau=10^4$. In short, the W-F coupling will take place once a
statistical equilibrium localized around the resonant frequency is
realized.

\section{Concluding remarks}

To study the time scale of the W-F coupling, we need a better
solution of the integro-differential equation of the resonant
scattering of Ly$\alpha$ photons. Especially, we need to know the
evolution of photon frequency distribution around the resonant
scattering. That is, the algorithm should be able to handle the
extremely flat distribution and its sharp boundary. These features
can properly be captured by the WENO scheme, which has high order of
accuracy and good convergence in capturing discontinuities as well
as to be significantly superior over piecewise smooth solutions
containing discontinuities. The WENO algorithm can be used for
resonant scattering with and without the recoil of atoms. This
algorithm is reliable as it passed all the tests.

The evolution of the photon distribution in the frequency space
generally undergoes three phases. In the first phase, the profile of
the photon distribution is similar to the initial one. In the second
phase, an extremely flat plateau (without recoil) or Boltzmann
distribution (with recoil) formed around the resonant frequency. The
width and height of the flat plateau or local Boltzmann distribution
increase with time. Finally, in the third phase, the photons
injected from the source is balanced by the redshift of the
expansion, and the evolution of the photon distribution is stable.
The first phase is very short.  The second phase will be onset after
a few tens or hundreds of photons scattering by atoms.  On the other
hand, the onset of the third phase is mainly dependent on the
Gunn-Peterson optical depth, which is large at early universe.
Consequently, the onset of the third phase is much later than the
second one.

Usually, the W-F coupling is described by time-independent solutions
of the Fokker-Planck approximation of the integro-differential
equation of the resonant scattering. With the WENO solutions, we
show that the time-independent solutions would not be available for
the 21 cm signal of the first generation stars, if the life time of
the evolution of the 21 cm region is equal to or less than 1 Myr.
However, the time scale of the onset of the W-F coupling actually is
irrelevant with the time-independent solutions. The W-F coupling
will take place once a statistical equilibrium is locally realized
in the frequency space around the resonant frequency. This time
scale is always a few tens or hundreds of the mean free flight time
of the resonant photons, and is generally independent of the
expansion of the universe when the Gunn-Peterson optical depth is
large, or the Sobolev parameter is much less than 1. More detail
results of the time-dependence o the W-F coupling is reported in Roy
et al (2009).

\noindent{\bf Acknowledgments}

This work is supported by the US NSF under the grants AST-0506734
and AST-0507340. We thank Dr. Jiren Liu for his help.
\appendix

\section{Numerical integration: an $\mathcal{O}(N)$ algorithm}

We need to numerically integrate $\displaystyle\int \mathcal{R}(x, x') J(x', t) dx'$, denoted as
\beq
I(x) =  \frac12 \int  e^{2bx' + b^2} {\rm erfc} (\max(|x+b|, |x'+b|)) J(x', t) dx',
\eeq
with $\mathcal{R}(x, x')$ as in eq.(\ref{eq7}).
To evaluate $I_m = I(x_m)$,  $\forall m=-N_l, \cdots, N_r$, we apply the rectangular rule,
which is spectrally accurate for smooth functions vanishing at boundaries,
\beqa
I_m
& =&  \frac 12 \int_{x_{left}}^{x_{right}} {\rm erfc} (\max(|x_m+b|, |x'+b|))e^{2bx' + b^2} J(x', t)dx' \\
&\approx&\frac12 \Delta x \sum_{i=-N_l}^{N_r} {\rm erfc} (\max(|x_m+b|, |x_i+b|))e^{2bx_i + b^2}J(x_i, t) .
\eeqa
Notice that this summation algorithm is very costly as it takes $\mathcal{O}(N)$
operations per $m$, therefore the total procedure has $\mathcal{O}(N^2)$ operations overall.
We use a grouping technique, described below, so that the overall computational cost
can be reduced to $\mathcal{O}(N)$, without changing mathematically the algorithm and its accuracy.

 The proposed scheme with order $N$ computational effort is the following.
Let $N_b = floor(\frac{b}{\Delta x})$ and $N_{2b} = floor(\frac{2b}{\Delta x})$.
The integration algorithm is designed for two cases:
$m \ge -N_b$ and $m< -N_b$.

In the case of $m \ge -N_b$ or equivalently $x_m + b \ge 0$:
\beqa
I_m  &=& \frac12 \Delta x  (\sum_{i=-N_l}^{-m-N_{2b}-1} {\rm erfc} (|x_i+b|)e^{2bx_i + b^2}J(x_i,
t)  \nonumber \\
& & +  {\rm erfc} (|x_m+b|) \sum_{i=-m - N_{2b}}^{m}e^{2bx_i + b^2}J(x_i, t) \nonumber\\
& & +\sum_{i=m+1}^{N_r} {\rm erfc} (|x_i+b|)e^{2bx_i + b^2}J(x_i, t))\\
&\doteq& \frac12 \Delta x (I_{1, m} + {\rm erfc} (|x_m+b|) I_{2, m} + I_{3,m})
\eeqa

\begin{enumerate}
\item Evaluate $I_{1, -N_b}$, $I_{2, -N_b}$ and $I_{3,-N_b}$ respectively as
\beqa
I_{1, -N_b} & = & \sum_{i=-N_l}^{N_b-N_{2b}-1} {\rm erfc} (|x_i+b|)e^{2bx_i + b^2}J(x_i, t), \\
I_{2, -N_b} &=& \sum_{i=N_b - N_{2b}}^{-N_b} e^{2bx_i + b^2}J(x_i, t), \\
I_{3,-N_b} &=& \sum_{i=-N_b+1}^{N_r} {\rm erfc} (|x_i+b|)e^{2bx_i + b^2}J(x_i, t),
\eeqa
which leads to $\mathcal{O}(N)$ cost.
\item
DO $m = -N_b+1, N_r$
Evaluate $I_{2, m}$, $I_{3, m}$ respectively by
\beq
I_{1, m} = I_{1, m-1} - {\rm erfc} (|x_{-m-N_{2b}}+b|) e^{2bx_{-m-N_{2b}}+ b^2}J(x_{-m-N_{2b}}, t)
\eeq
\beq
I_{2, m} = I_{2, m-1} + e^{2bx_{m} + b^2}J(x_{m}, t) + e^{2bx_{-m-N_{2b}} + b^2}J(x_{-m-N_{2b}}, t)
\eeq
\beq
I_{3, m} = I_{3, m-1} -{\rm erfc} (|x_{m}+b|)e^{2bx_{m} + b^2}J(x_{m}, t)
\eeq
ENDDO

To be consistent with the indeces, if $N_l - N_{2b} < N_r$ then, we will set $I_{1, m} = 0$, for $m = N_l - N_{2b}, N_r$.
The algorithm leads to $\mathcal{O}(1)$ cost per $m$, therefore $\mathcal{O}(N)$ computation overall.
\end{enumerate}

In the case of $m < -N_b$, or equivalently $x_m + b < 0$:
\beqa
I_m  &=& \frac12 \Delta x  (\sum_{i=-N_l}^{m-1} {\rm erfc} (|x_i+b|)e^{2bx_i + b^2}J(x_i, t)  \nonumber\\
&&+  {\rm erfc} (|x_m+b|) \sum_{i=m}^{-m-N_{2b}-1}e^{2bx_i + b^2}J(x_i, t) \nonumber\\
& & +\sum_{i=-m-N_{2b}}^{N_r} {\rm erfc} (|x_i+b|)e^{2bx_i + b^2}J(x_i, t))\\
&=& \frac12 \Delta x (I_{1, m} + {\rm erfc} (|x_m+b|) I_{2, m} + I_{3,m})
\eeqa
\begin{enumerate}
\item Evaluate $I_{1, -N_b-1}$, $I_{2, -N_b-1}$ and $I_{3,-N_b-1}$ as
\beqa
I_{1, -N_b-1} &=& \sum_{i=-N_l}^{-N_b-2} {\rm erfc} (|x_i+b|)e^{2bx_i + b^2}J(x_i, t), \\
I_{2, -N_b-1} &=&\sum^{N_b - N_{2b}}_{i=-N_b-1} e^{2bx_i + b^2}J(x_i, t), \\
I_{3,-N_b-1} &=&\sum_{i=N_b+1-N_{2b}}^{N_r} {\rm erfc} (|x_i+b|)e^{2bx_i + b^2}J(x_i, t),
\eeqa
which leads to $\mathcal{O}(N)$ cost.
\item
DO $m = -N_b-2, -N_l$

Evaluate $I_{1, m}$, $I_{2, m}$, $I_{3, m}$ respectively by
\beq
I_{1, m} = I_{1, m+1} - {\rm erfc} (|x_{m}+b|) e^{2bx_{m}+ b^2}J(x_{m}, t)
\eeq
\beq
I_{2, m} = I_{2, m+1} + e^{2bx_{m} + b^2}J(x_{m}, t) + e^{2bx_{-m-N_{2b}-1} + b^2}J(x_{-m-N_{2b}-1}, t)
\eeq
\beq
I_{3, m} = I_{3, m+1} - {\rm erfc} (|x_{-m-N_{2b}-1}+b|)e^{2bx_{-m-N_{2b}-1} + b^2}J(x_{-m-N_{2b}-1}
, t)
\eeq
ENDDO

To be consistent with the indeces, if $N_l - N_{2b} > N_r$, we will set $I_{3, m} = 0$, for $m= -N_{2b} - N_r - 1, -N_l$.
Again, the algorithm leads to $\mathcal{O}(1)$ cost per $m$, therefore $\mathcal{O}(N)$ computation overall.
\end{enumerate}

\end{document}